\documentclass[prl,twocolumn,floatfix,nopacs]{revtex4-1}
\usepackage{graphicx,amsfonts,amssymb,amsmath,hyperref,enumerate,upgreek,tabularx,multirow}
\usepackage[displaymath,mathlines]{lineno}

\newif\ifhyper
\hypertrue
\ifhyper
\hypersetup{
   citecolor = {green},
   colorlinks = {true}, 
   urlcolor = {blue} 
}
\fi

\newcommand{\beq}{\begin{equation}}
\newcommand{\eeq}{\end{equation}}
\newcommand{\beqa}{\begin{eqnarray}}
\newcommand{\eeqa}{\end{eqnarray}}

\newcommand{\subfig}[2]{%
    {\large \textsf{\textbf{#1}}} \vtop{%
  \vskip0pt
  \hbox{#2}
}}

\def\Longarrow{\protect\@lra}
\def\@lra{\relbar\joinrel\relbar\joinrel\relbar\joinrel%
          \relbar\joinrel\rightarrow}

\begin{document}


\title{An Operational Definition of Topological Order}

\author{Amit Jamadagni}
\email{amit.jamadagni@itp.uni-hannover.de}
\affiliation{Institut f\"ur Theoretische Physik, Leibniz Universit\"at Hannover, Appelstra{\ss}e 2, 30167 Hannover, Germany}
\author{Hendrik Weimer}
\affiliation{Institut f\"ur Theoretische Physik, Leibniz Universit\"at Hannover, Appelstra{\ss}e 2, 30167 Hannover, Germany}

\begin{abstract}

  The unrivaled robustness of topologically ordered states of matter
  against perturbations has immediate applications in quantum
  computing and quantum metrology, yet their very existence poses a
  challenge to our understanding of phase transitions. However, a
  comprehensive understanding of what actually constitutes topological
  order is still lacking. Here we show that one can interpret
  topological order as the ability of a system to perform topological
  error correction. We find that this operational approach
  corresponding to a measurable both lays the conceptual foundations
  for previous classifications of topological order and also leads to
  a successful classification in the hitherto inaccessible case of
  topological order in open quantum systems. We demonstrate the
  existence of topological order in open systems and their phase
  transitions to topologically trivial states. Our results demonstrate
  the viability of topological order in nonequilibrium quantum systems
  and thus substantially broaden the scope of possible technological
  applications.

\end{abstract}

\maketitle

\section{Introduction}

Topologically ordered phases are states of matter that fall outside of
Landau's spontaneous symmetry breaking paradigm and cannot be
characterized in terms of local order parameters, as it is the case
with conventional symmetry-breaking phase transitions.  From a broad
perspective, they can be classified into symmetry-protected
topological order or intrinsic topological order \cite{Wen2017}.  For
the former, the existence of a symmetry is required to maintain
topological order, i.e., when the symmetry is broken the system
immediately returns to a topologically trivial state. In some cases,
topological order can be captured in terms of topological invariants
such as the Chern number \cite{Thouless1982,Kohmoto1985}, but being
based on single-particle wave functions, their extension to
interacting systems is inherently difficult
\cite{Fidkowski2010}. Alternatively, topological order has been
discussed in terms of nonlocal order parameters often related to
string order \cite{denNijs1989,Haegeman2012,Pollmann2012,Elben2020},
but the main difficulty of this approach is that such string order can
also be observed in topologically trivial phases
\cite{DallaTorre2006}. From a conceptual point of view, a particularly
attractive definition of topological order is the impossibility to
create a certain quantum state from a product state by a quantum
circuit of finite depth \cite{Chen2010}. However, since this is
equivalent to the uncomputable quantum Kolmogorov complexity
\cite{Mora2007}, it has very little practical applications. Hence,
most analyses of topological ordered systems have been centered around
indirect signatures such as the topological entanglement entropy
\cite{Preskill2006,Levin2006,Balents2012} or minimally entangled
states \cite{Zhang2012,Zhu2013}, but even those quantities can prove
difficult to interpret \cite{Bridgeman2016,Jamadagni2018}.

Here, we overcome the limitations of the previous approaches to
topological order by understanding topological order as the intrinsic
ability of a system to perform topological error correction, giving
rise to an operational definition of topological order that can be
readily computed. Our definition is thus connected to
  taking the robustness of topological phases as its defining property
  \cite{Nussinov2009,Qiu2020}, see the Supplementary Methods for a
  detailed technical discussion. To make our definition
mathematically precise, we call a system to be in a topologically
ordered state if it can be successfully corrected by an error
correction circuit of finite depth. One key advantage of our approach
is that the error correction circuit does not have to be optimal, as
it only requires to reproduce the correct finite size scaling
properties, which can be expected to be universal across a topological
phase transition. This puts our approach in stark contrast with the
classification of topological error correction codes in terms of their
threshold values \cite{Terhal2015}, as the latter is a nonuniversal
quantity. Symmetry protected topological order can be represented
within our error correction formalism by imposing certain symmetry
constraints on the error correction circuit. Compared to previous
approaches to topological order, another striking advantage of our
error correction method is that it corresponds to an actual
observable, which can be measured in an experiment.

\begin{figure*}[t]
\begin{center}
 \begin{tabular}{lp{0.25cm}c}
    \subfig{a}{\hspace{0.25cm}\includegraphics[width=.45\linewidth]{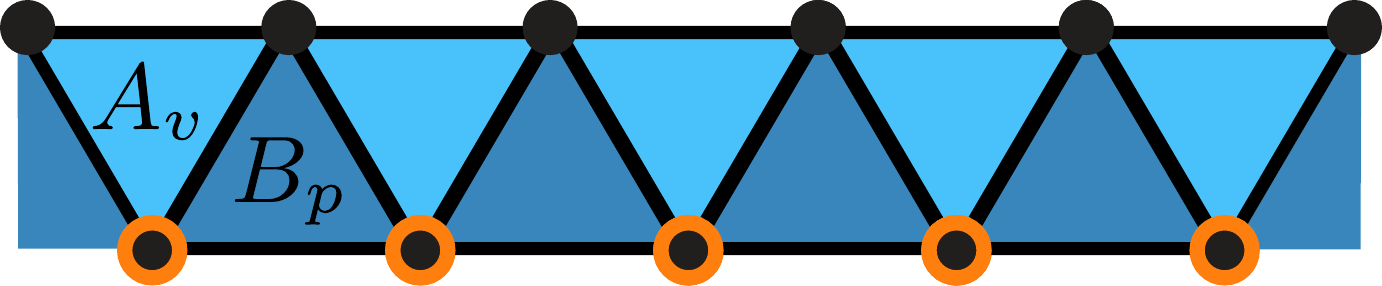}}\vspace{0.25cm}
    &&
    \multirow{2}{*}{\subfig{c}{\includegraphics[width=.45\linewidth]{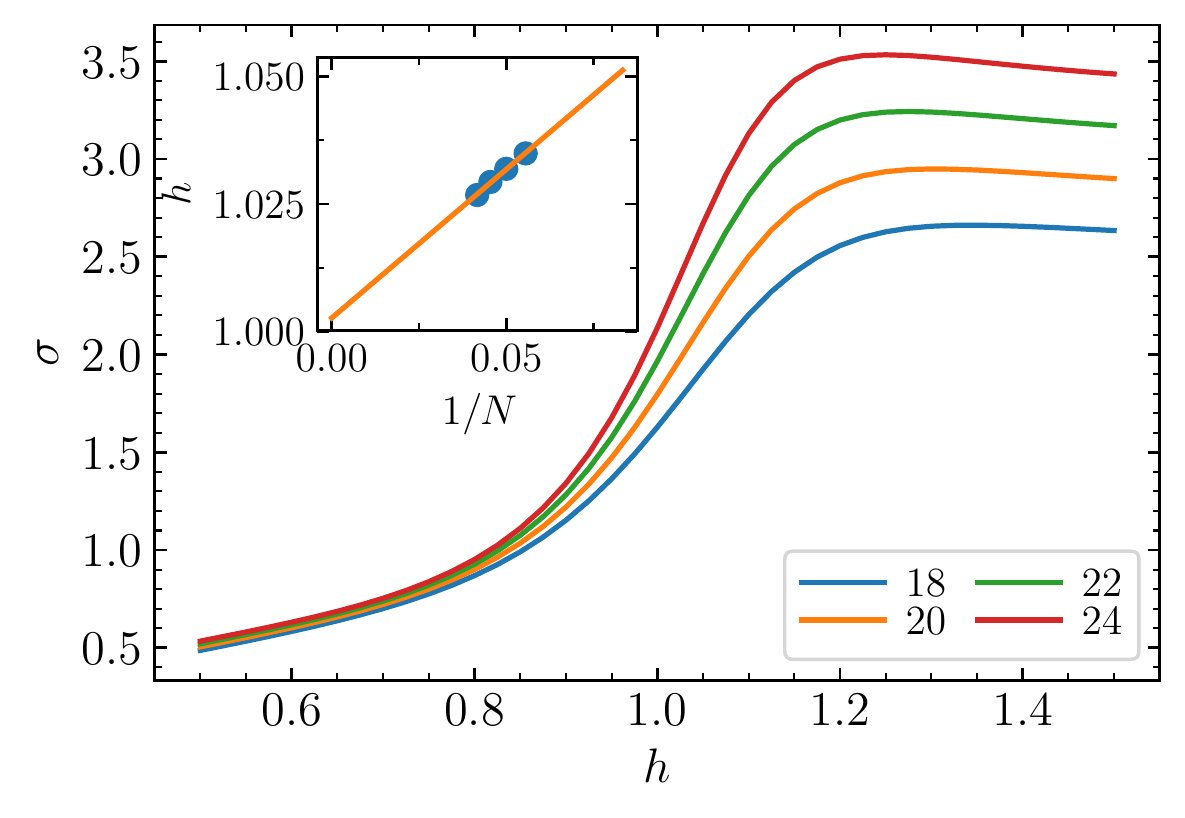}}}\\
    \subfig{b}{\hspace{0.25cm}\includegraphics[width=.45\linewidth]{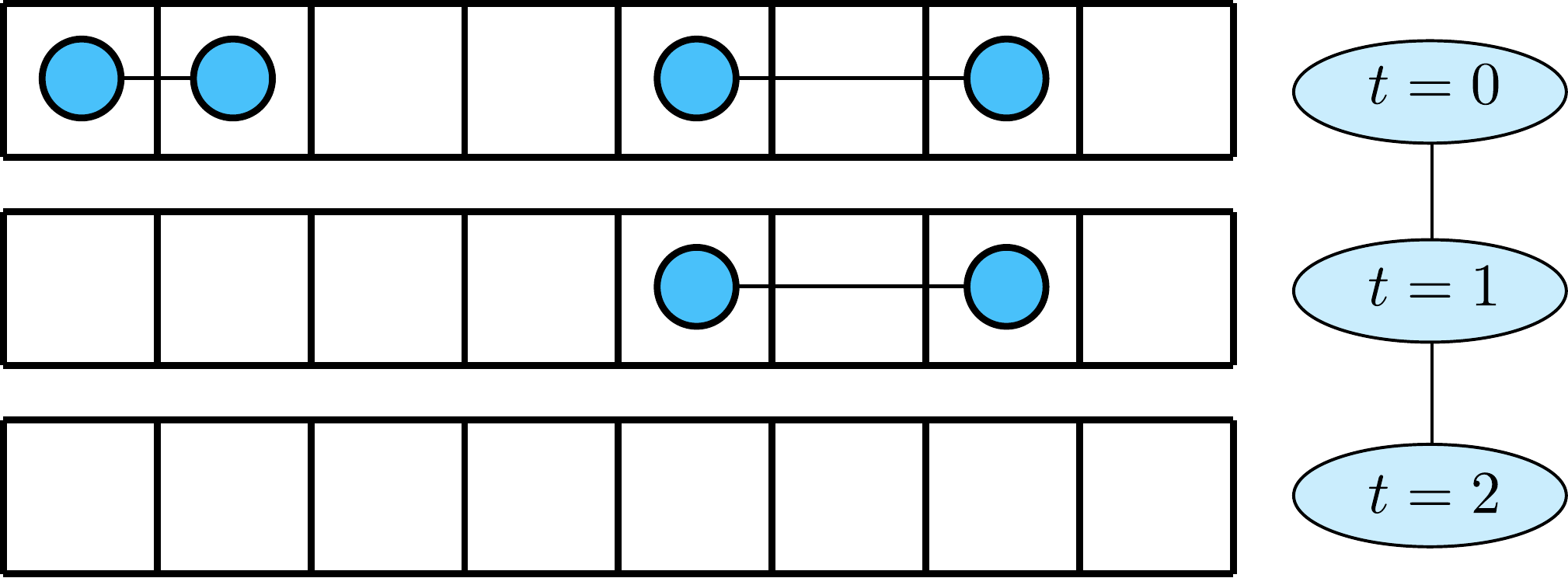}}
  \end{tabular}
\end{center}

\caption{Topological order in a quasi-1D toric code model. (a) The
  $A_v $ and $B_p$ operators are arranged along two rails of sites
  such that the perturbation on the lower rail (orange) maps onto the
  1D transverse field Ising model. (b) Example of the error correction
  procedure for $N=8$ Ising spins containing four errors, with the
  state of the system shown after $t$ timesteps of the algorithm. The
  errors are fused along the horizontal strings, with the total error
  correction depth being $t_d = 2$. (c) Standard deviation $n_\sigma$
  of the circuit depth for different system sizes. Above the
  topological transition, the circuit depth diverges in the
  thermodynamic limit, with a finite size scaling analysis (inset)
  yielding critical value of $h_c = 1.003(1)$.}
  \label{fig:ising}
\end{figure*}

\section{Results}
\subsection{Operational definition for closed systems}
To demonstrate the viability of our error correction approach in a
concrete setting, we turn to the toric code model, which serves as a
paradigm for intrinsic topological order \cite{Kitaev2003}. Its
Hamiltonian is given by a sum over two classes of spin $1/2$ operators
describing four-body interactions, $A_v$ and $B_p$, acting on
vertices $v$ and plaquettes $p$, respectively, according to
\begin{linenomath}
\begin{equation}
  H_{TC} = -E_0\left(\sum\limits_v \underbrace{\sigma_\alpha^{x}\sigma_\beta^{x}\sigma_\gamma^{x}\sigma_\delta^{x}}_{A_v} + \sum\limits_p \underbrace{\sigma_\mu^{z}\sigma_\nu^{z}\sigma_\rho^{z}\sigma_\sigma^{z}}_{B_p}\right),
  \label{eq:toric}
\end{equation}
\end{linenomath}
where $\sigma_i^{x,z}$ denotes the Pauli matrix acting on site
$i$. The robustness of topological order of the ground state can be
analyzed with respect to the response of a perturbation describing a
magnetic field, i.e., $H=H_{TC} - h\sum_i\sigma_i^x$. Importantly, the
perturbed toric code can be mapped onto an Ising model in a transverse
field using a highly nonlocal unitary transformation
\cite{Tagliacozzo2011}. The phase transition from the topologically
ordered to the trivial state then corresponds to the phase transition
between the paramagnet and the ferromagnet in the Ising model
\cite{Trebst2007}. Here, we will be interested in the case where the
perturbed toric code can be mapped exactly onto the one-dimensional
(1D) Ising model, which can be realized by imposing the right boundary
condition \cite{Jamadagni2018}, see Fig.~\ref{fig:ising}. Our approach
has the advantage that the critical point of the topological phase
transitions is known to be exactly at $h_c = 1$. As
  there is no intrinsic topological order in 1D systems, the phase for
  $h<1$ is actually a symmetry-protected topological phase, with the
  protected symmetry being the $Z_2$ symmetry of the associated Ising
  model.
  Additionally, note that the quasi-1D nature of the toric
code model results in the four-body interactions in
Eq.~(\ref{eq:toric}) being replaced by three-body interactions.

For the topological order arising in the toric code, the required
topological error correction can be readily expressed in terms of the
Ising variables $S_v = A_v$ and $S_p = B_p$, where each spin having
$S_i^z = -1$ corresponds to the presence of an error. As a first step
of the error correction algorithm, a syndrome measurement is
performed, i.e., all the Ising spins are measured in their $S^z$
basis, corresponding to the measurement of both the $A_v$ and $B_p$
degrees of freedom in the original toric code model. Under the
perturbation, the observables $S_i^z$ exhibit quantum fluctuations,
therefore it is necessary to perform a statistical interpretation of
the depth of the error correction circuit. Here, we find that the
standard deviation of the circuit depth exhibits substantially better
finite size scaling behavior than the mean, hence we use the former
for the detection of topological order in the following. The error
correction circuit is then implemented in a massively parallel way by
decorating each of the detected errors by a walker that travels
through the system until it encounters another error, upon which the
two errors are fused and removed from the system \cite{Wootton2015},
see the Methods section for details. Figure \ref{fig:ising}
demonstrates that our error correction approach is indeed able to
detect the topological phase transition, including the identification
of the correct critical point at $h_c = 1$.
Crucially,
we want to stress that our notion of error correction is not
limited to toric code models. In the Supplementary Discussion, we
consider the cubic code model \cite{Haah2011} hosting fracton
topological order \cite{Chamon2005}.

\begin{figure*}[t]
\begin{center}
  \begin{tabular}{cp{0.5cm}c}
    \subfig{a}{\includegraphics[width=.45\linewidth]{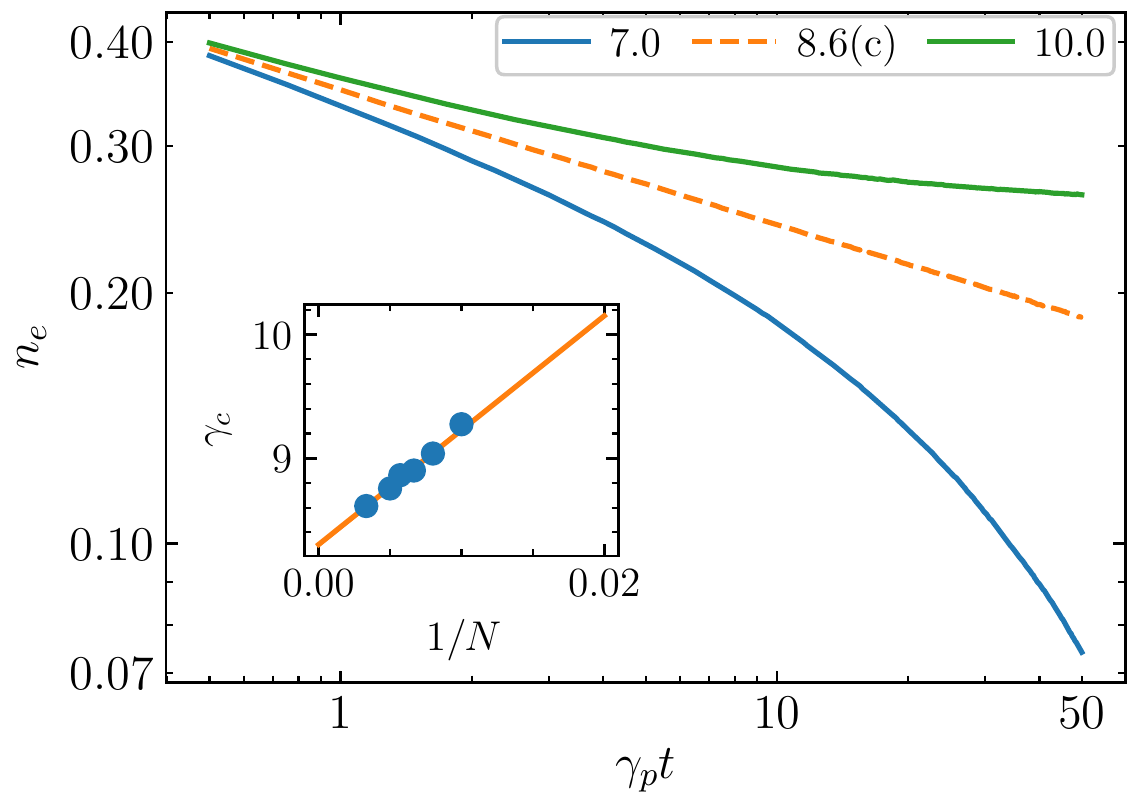}}
    &&
    \subfig{b}{\includegraphics[width=.45\linewidth]{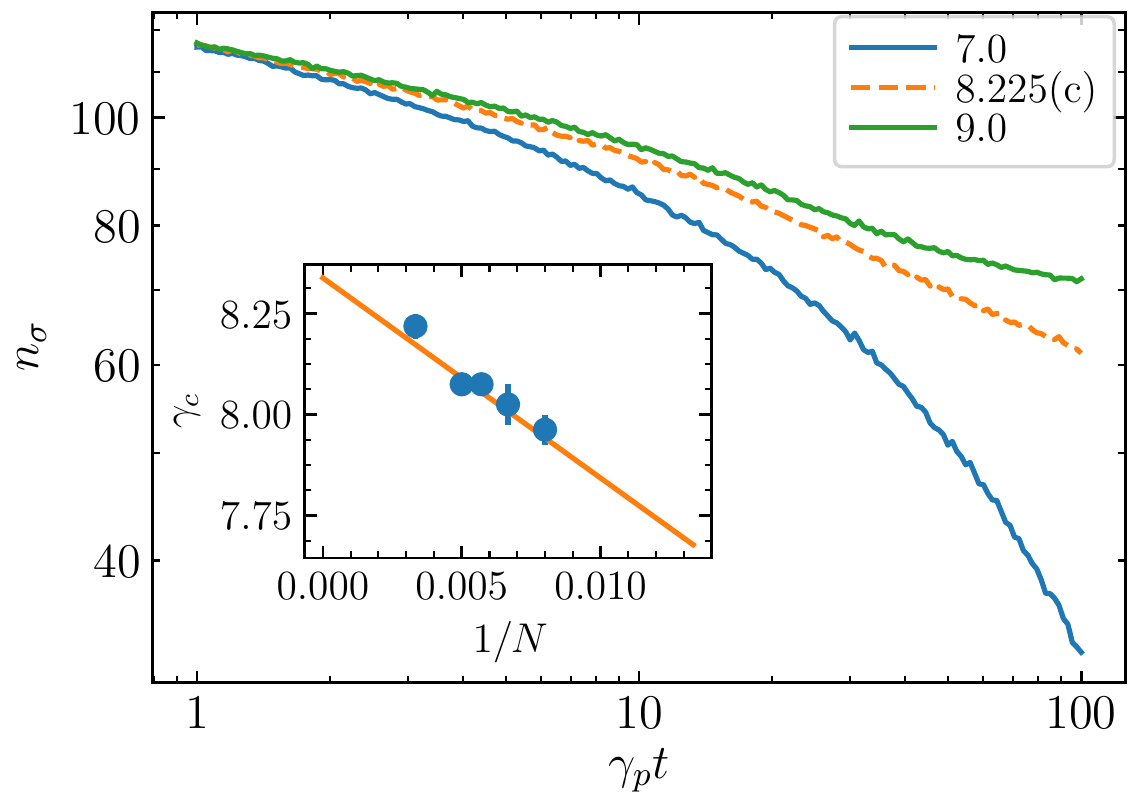}}
  \end{tabular}
\end{center}
\caption{Topological absorbing state transition. Error density $n_e$
  (a) and circuit depth $n_\sigma$ (b) for $N=300$ sites and different
  values of $\gamma$, showing subcritical behavior (blue), critical
  behavior (orange), and supercritical behavior (green). Initial
  states were chosen to have maximum $n_e$ or $n_\sigma$,
  respectively.  Finite size scaling leads to $\gamma_c = 8.30(2)$ (a,
  inset) and $\gamma_c = 8.34(5)$ (b, inset) in the thermodynamic limit.}

  \label{fig:minst}

\end{figure*}

\subsection{Topological order in open quantum systems}
\begin{figure*}[t]
\begin{center}
  \begin{tabular}{cp{0.5cm}c}
    \subfig{a}{\includegraphics[width=.45\linewidth]{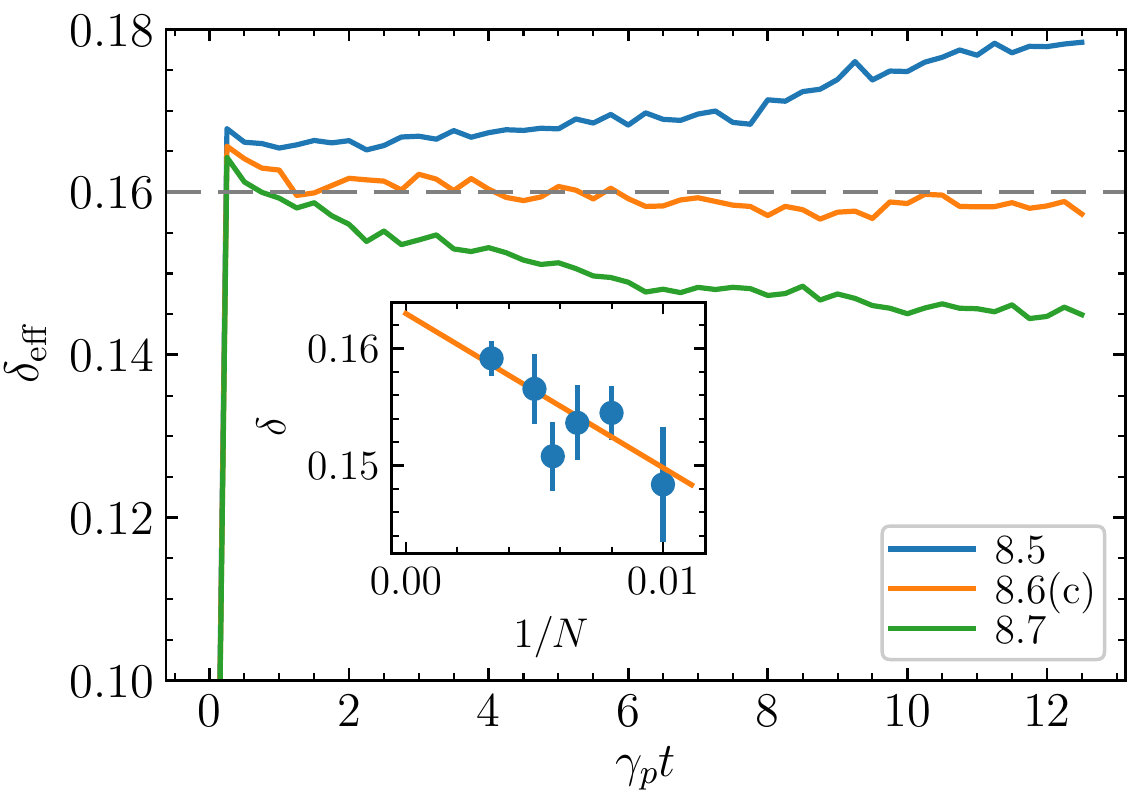}}
    &&
    \subfig{b}{\includegraphics[width=.45\linewidth]{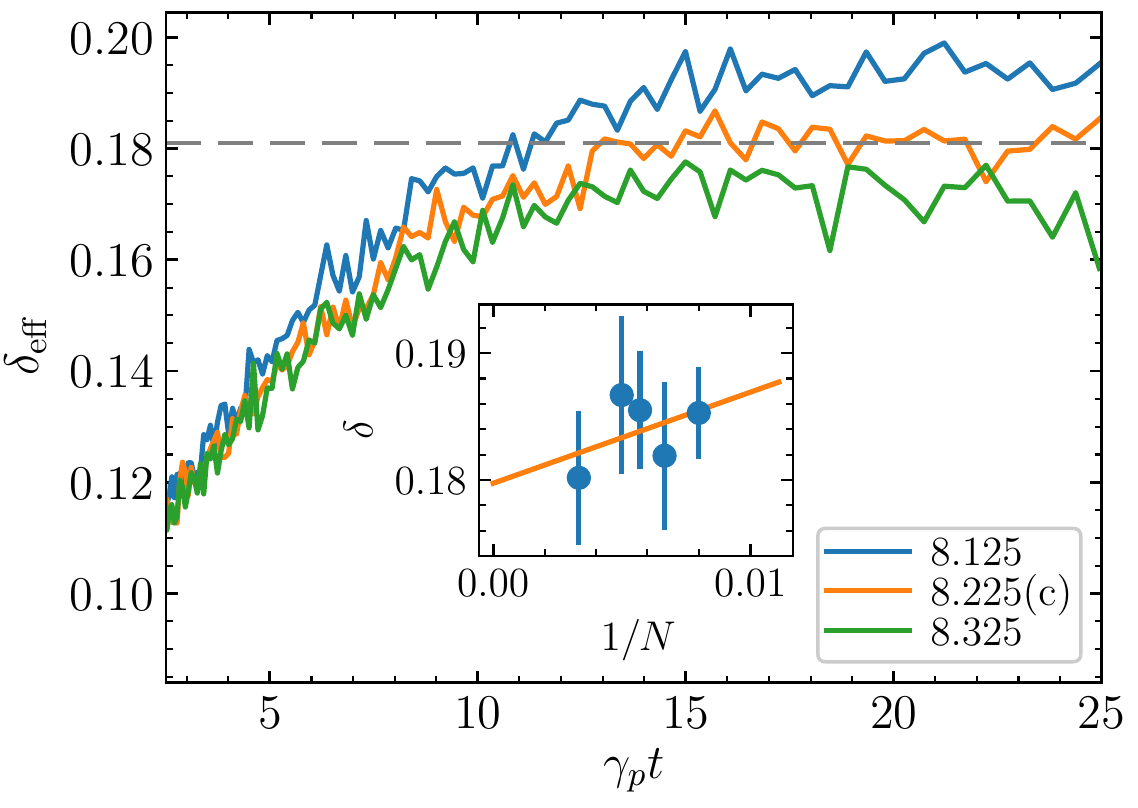}}
  \end{tabular}
\end{center}
\caption{Effective critical exponents according to
  Eq.~(\ref{eq:delta}) for the absorbing-active transition (a) and the
  topological transition (b) ($m=4$). The critical value of the
  transition is taken where $\delta_\text{eff}$ remains
  constant. Error bars correspond to all values consistent with a
  constant value in the long time limit. Finite size scaling leads to
  $\delta = 0.163(5)$ (a, inset) and $\delta = 0.18(2)$ (b, inset) in
  the thermodynamic limit. Errors are given by the sum of the
  uncertainty in the linear fit and the difference in $\delta$ between
  $m=4$ and $m=2$.}

  \label{fig:delta}

\end{figure*}

Let us now extend our approach to mixed quantum states, where previous
works have shed some light on topological properties
\cite{Hastings2011,Bardyn2013,Viyuela2014,Huang2014,Grusdt2017,Roberts2017},
but a universally applicable definition of topological order has
remained elusive so far. This extension is straightforward, as the
implementation of the topological error correction channel can be
applied to mixed states as well. Here, we consider mixed states
arising in open quantum systems with purely dissipative dynamics given
in terms of jump operators $c_i$ according to the Markovian quantum
master equation $d\rho/dt = \sum_i c_i\rho
c_i^{\dagger}-\{c_i^{\dagger} c_i,\rho\}/2$. Dissipative variants of
the toric code can be constructed by considering the jump operators
\begin{linenomath}
  \begin{align}
    \label{eq:cool1}
  c_i^v &= \sqrt{\gamma_v}\sigma_i^z(1-A_v)/2, i\in v\\
   c_i^p &= \sqrt{\gamma_p}\sigma_j^x(1-B_p)/2, j\in p
  \label{eq:cool2}
\end{align}
\end{linenomath}
with rates $\gamma_{v,p}$, which result in the toric code ground
states being steady states of the quantum master equation
\cite{Weimer2010}. As before, we now consider the robustness of
topological order to an additional perturbation. Here, we will first
consider again a quasi one-dimensional model analogous to
Fig.~\ref{fig:ising}a , in which the perturbation is given by
\begin{linenomath}
\begin{equation*}
  c_i^h = \sqrt{\gamma}\sigma_i^x(1-B_p)/2, i \in p+1,
\end{equation*}
\end{linenomath}
with $i$ being restricted to the upper rail. Note that
  in contrast to the jump operator of Eq.~(\ref{eq:cool1}), this jump
  operator involves a spin that is part of the plaquette $p+1$ and not of
  the plaquette $p$. This multi-plaquette operator leads to a heating
  process introducing new errors, while the jump operators of
  Eqs.~(\ref{eq:cool1}, \ref{eq:cool2}) describe cooling processes that
  remove errors from the system. Importantly, the creation of a new
error on the plaquette $p+1$ requires the existence of another error
on the neighboring plaquette $p$, which is also reflected after
mapping onto Ising variables, see the Methods section for
details. This results in the model falling into the well-known class
of absorbing state models \cite{Hinrichsen2000}, with the toric code
ground state corresponding to the absorbing state. Such absorbing
state models can exhibit phase transitions to an active phase where
the absorbing state is no longer reached asymptotically when starting
from a different initial state. Here, we indeed find such a phase
transition in the density of errors, see
Fig.~\ref{fig:minst}. Moreover, this absorbing-to-active transition is
also accompanied by a divergence of the depth of the error correction
circuit, i.e., by a topological transition to a trivial phase. We also
track the critical exponent $\delta$ measuring the algebraic decay of
the density of errors $n_e$ or the circuit depth $n_\sigma$,
respectively, by considering the quantity
\begin{equation}
  \delta_{\text{eff}}(t) = -\frac{1}{\log m}\log\frac{n_{e,\sigma}(mt)}{n_{e,\sigma}(t)}
  \label{eq:delta}
\end{equation}
which remains constant for a fixed value of $m$ \cite{Hinrichsen2000}. In
the limit of large system sizes, both the critical strength for the
transition and the critical exponent are in close agreement between
the absorbing-to-active transition and the topological transition, see
Fig.~\ref{fig:delta}, belonging to the universality class
of one-dimensional directed percolation ($\delta =
0.163$\cite{Hinrichsen2000}).

\begin{figure*}[t]
\begin{center}
  \begin{tabular}{cp{0.5cm}c}

    \subfig{a}{\includegraphics[width=.45\linewidth] {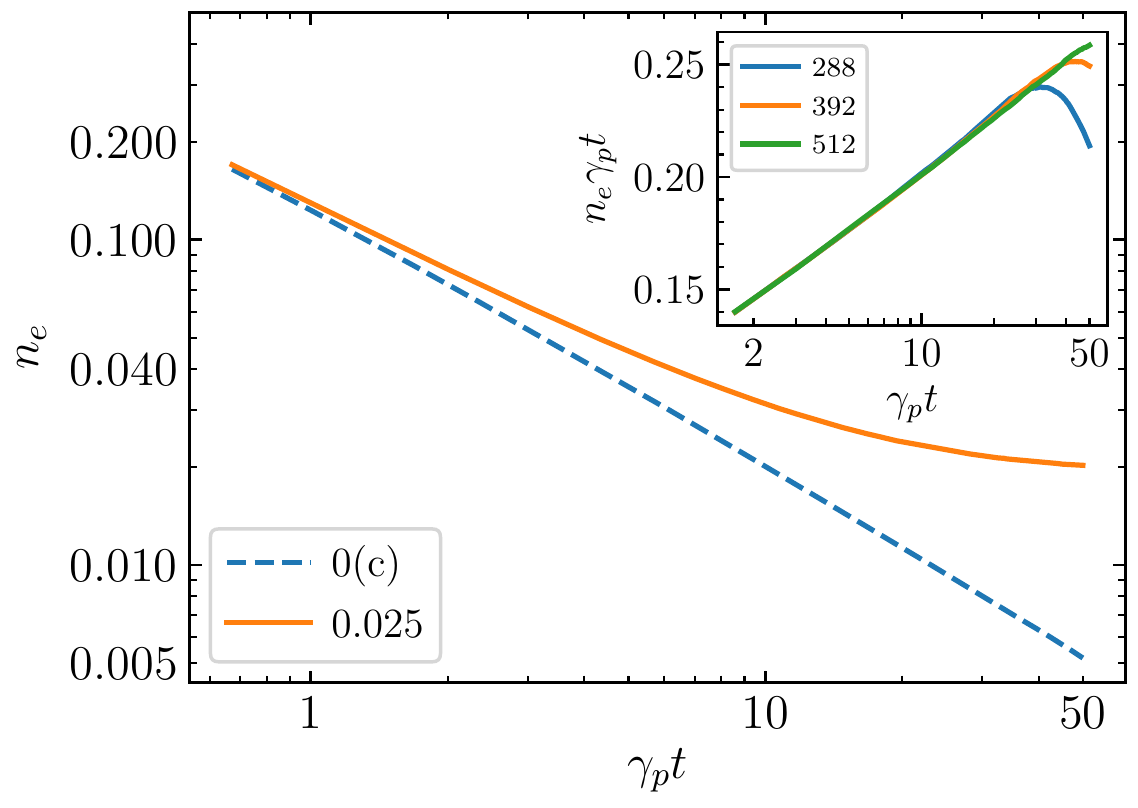}}
    &&
    \subfig{b}{\includegraphics[width=.45\linewidth] {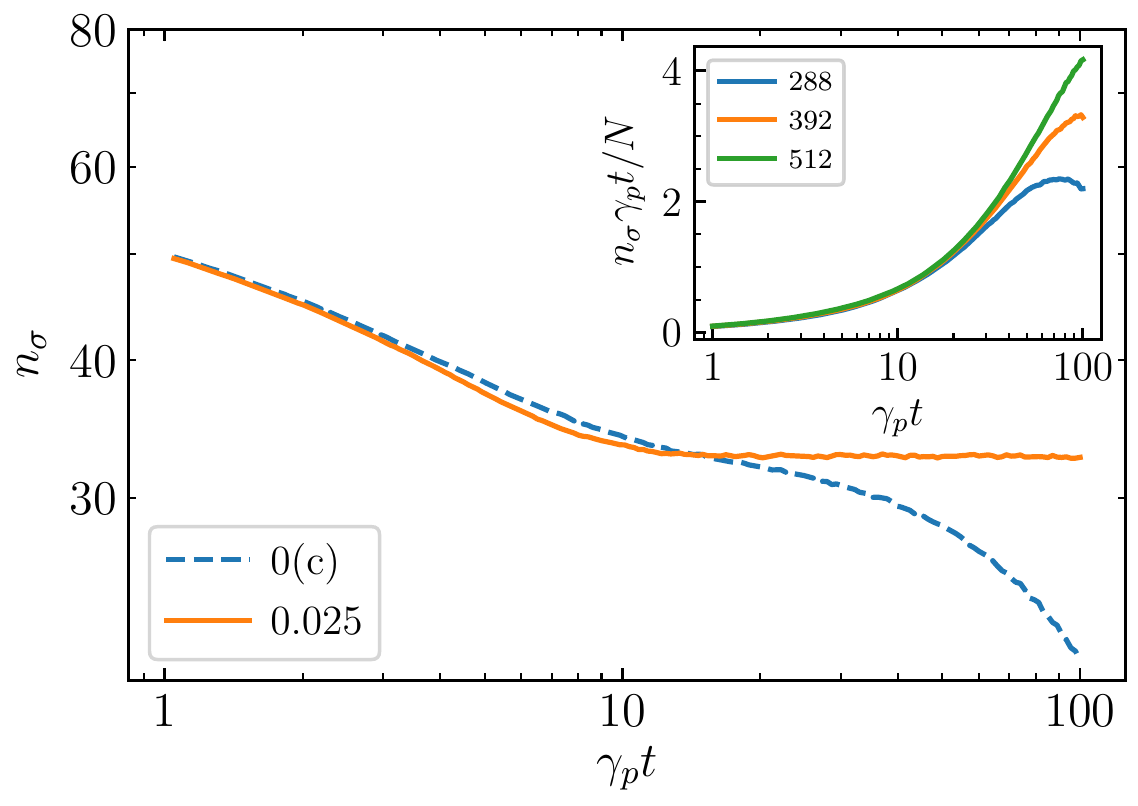}}
  \end{tabular}
\end{center}
\caption{Two-dimensional topological criticality. Error density $n_e$
  (a) and circuit depth $n_\sigma$ (b) for $N=512$ sites and $\gamma =
  0$ (dashed) and $\gamma = 0.025\,\gamma_p$ ($\gamma_p =
  \gamma_v$). The insets show the logarithmic corrections to a
  $t^{-1}$ decay, with a linear behavior for the error density (a) and
  a quadratic behavior for the topological transition (b) before
  finite size effects become relevant.}
\label{fig:barw}
\end{figure*}

Importantly, our approach to topological order can also be readily
applied to higher-dimensional systems. Here, we will be interested in
a two-dimensional absorbing state model, in which both error types are
present. In particular, the creation of $A_v$ errors is conditional on
the existence of a neighboring $B_p$ error and vice versa.  Hence, we
consider jump operators of the form
\begin{linenomath}
\begin{align*}
  c_i^{hv} &= \sqrt{\gamma}\sigma_i^x(1-A_v)/2\\
  c_i^{hp} &= \sqrt{\gamma}\sigma_i^z(1-B_p)/2.
\end{align*}
\end{linenomath}
Importantly, the lack of boundary processes now leads to a
conservation of the parity of both type of errors. This model can be
expected to be in the same universality class as two-dimensional
branching-annihilating random walks with two species
\cite{Odor2001}. While the model is active for any finite $\gamma$,
it exhibits nontrivial critical behavior, having an exponent $\delta =
1$ with logarithmic corrections. Figure \ref{fig:barw} shows the data
collapse for different system sizes for both the error density and the
circuit depth, confirming this picture. Strikingly, the logarithmic
corrections in the topological case include a quadratic term that is
not present in the error density, pointing to a different critical
behavior. This demonstrates that topological criticality cannot be
predicted using only the properties of an accompanying conventional
phase transition.

\subsection{Summary}
In summary, we have introduced a novel operational approach to
topological order based on the ability to perform topological error
correction. Our method reproduces known topological phase transitions
and can be readily applied to previously inaccessible cases such as
topological transitions in open quantum systems, and has the
additional advantage that it corresponds to a measurable
observable. Furthermore, we would like to note that our approach can
be readily applied to other topologically ordered systems. For
instance, both the Kitaev wire \cite{Kitaev2001} and Haldane
insulators \cite{Haldane1983} can be mapped onto effective spin
models, where an analogous error correction strategy can be performed.
Finally, our work paves the wave for many future theoretical and
experimental investigations, such as the application of our approach
to fracton \cite{Chamon2005,Haah2011,Brown2020} or Floquet
\cite{Kitagawa2010,Lindner2011} topological order, or the direct
experimental realization of the error correction protocol presented in
our work for the development of future quantum technological devices.

\section{Methods}

\subsection{Error correction in toric code models}

\begin{figure}[bt]
\begin{center}
  \begin{tabular}{l}
    \subfig{a}{\hspace{0.25cm}\includegraphics[width=.8\linewidth]{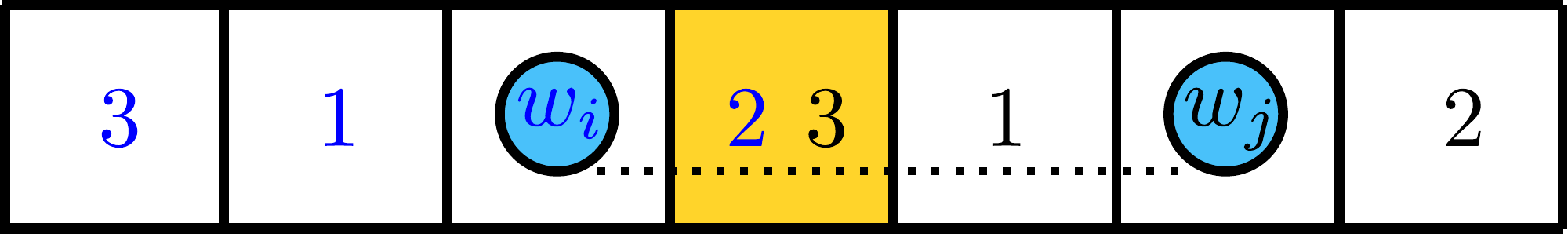}}\\[1.5cm]
    \subfig{b}{\hspace{0.25cm}\includegraphics[width=.8\linewidth]{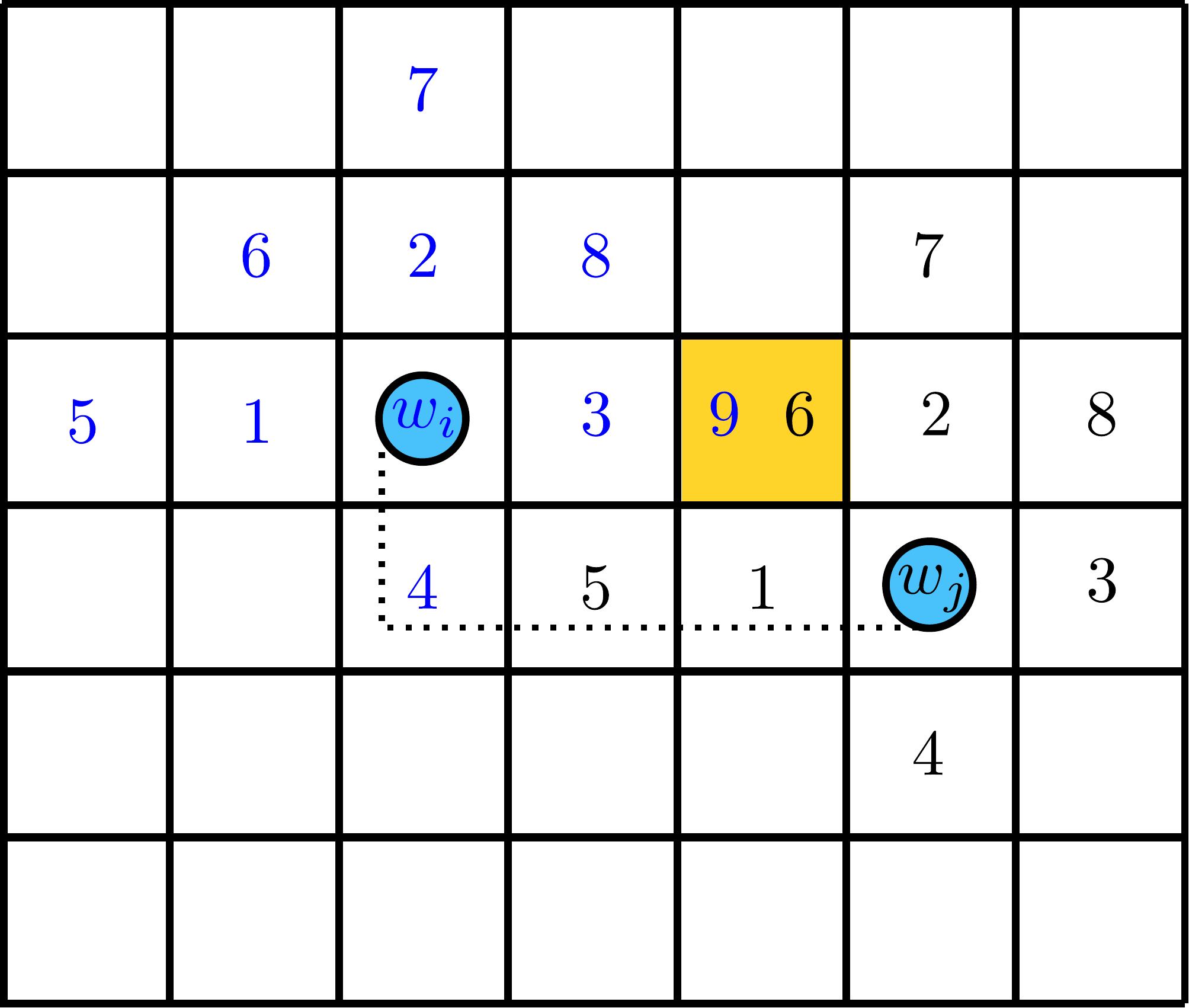}}
  \end{tabular}
\end{center}
  \caption{Schematic representation of the error correction
    procedure. Two errors are associated with walkers $w_i$ (blue) and
    $w_j$ (black), located at the errors at $t=0$. The colored numbers
    indicate the timestep at which a particular walker visits a
    site. Once a walker encounters a site already visited by the other
    walker (yellow), the two errors can be fused along the dotted
    path. In one dimension (a), the walkers alternate in a left-right
    pattern, in two dimensions (b), the walkers proceed in
    diamond-shaped patterns corresponding to a constant Manhattan
    distance from the initial sites.}

  \label{fig:walkers}
\end{figure}
The error correction scheme for the detection of topological order is
based on the results from the error syndrome measurements, which can be
cast in terms of the spin variables $S_v$ and $S_p$. In cases where
there are both error types being present, the error correction can be
realized independently. As the figure of merit, we are interested in
the depth of the classical error correction circuit, which maps the
initial erroneous state onto the topologically ordered state without
any errors. Our error correction procedure is massively parallelized,
i.e., within a topologically ordered phase, it is able to remove a
thermodynamically large number of errors in constant time. This is in
stark contrast to the conventional maximum-likelihood error correction
\cite{Dennis2002}, as this will always require an error correction
circuit whose depth scales with the system size. The same argument
also holds for assessing topological order based on circuit complexity
\cite{Liu2020}, as the circuit complexity is an extensive quantity
even in the topologically ordered phase.

For each error, we decorate the associated site with a walker $w_i$,
which continuously explores the surroundings of the original site,
looking for the presence of other errors. In one spatial dimension,
the walker alternates between investigating sites on the left and on
the right, while in two dimensions, this is generalized to
continuously exploring sites with an increasing Manhattan distance to
the original site, see Fig.~\ref{fig:walkers}. For
simplicity, we assume that changing the site of a walker takes exactly
one unit of time, irrespectively of the distance traveled. Once a
walker encounters a site with either an error or a site previously
visited by a walker $w_j$ originating from another error, the error
correction procedure starts. For this, the errors on site $i$ and $j$
are fused together along the shortest path, removing them and their
associated walkers from the system. Here, we assume that the fusion is
instantaneous, which does not modify the overall finite size scaling
properties of the error correction circuit. The error correction
procedure is performed until all errors have been removed from the
system. In the one-dimensional case, we also allow for errors being
removed via the left or right boundary of the system, preventing the
case of a single error remaining without a potential fusion partner.

\subsection{Ising-mapped jump operators}

After mapping the system onto Ising variables $S_i$, we obtain a
purely classical master equation, despite the basis states being
highly entangled. Here, we take the limit $\gamma_v \to \infty$ such
that the dynamics is restricted to the Ising spins related to the
$B_p$ operators. In the basis of the Ising spins $S_i$, we obtain the
jump operators
\begin{linenomath}
\begin{align*}
  c_i^p &= \sqrt{\gamma_p}S_i^xS_{i+1}^x(1-S_{i}^z)/2\\
  c_i^h &= \sqrt{\gamma}S_{i+1}^x(1-S_{i}^z)/2.
\end{align*}
\end{linenomath}

\section{Data availability}
  
The data that support the plots within this paper and other findings
of this study are available from the corresponding author upon
reasonable request.

\section{Code availability} 

All numerical codes in this paper are available upon request to 
the authors.

\section{Acknowledgments}

We thank S.~Diehl and T.~Osborne for fruitful discussions. This work
was funded by the Volkswagen Foundation, by the Deutsche
Forschungsgemeinschaft (DFG, German Research Foundation) within SFB
1227 (DQ-mat, project A04), SPP 1929 (GiRyd), and under Germany’s
Excellence Strategy -- EXC-2123 QuantumFrontiers -- 390837967.

  \section{Author contributions}

Both authors contributed extensively to all parts of the manuscript.

  \section{Competing interests}

  The authors declare no competing interests.

\bibliographystyle{nature}
\bibliography{topo}

\clearpage

\onecolumngrid

\appendix

\end{document}




%
%

\title{Supplementary Information for ``An operational definition of topological order''}

\author{Amit Jamadagni}
\affiliation{Institut f\"ur Theoretische Physik, Leibniz Universit\"at Hannover, Appelstra{\ss}e 2, 30167 Hannover, Germany}
\author{Hendrik Weimer}
\affiliation{Institut f\"ur Theoretische Physik, Leibniz Universit\"at Hannover, Appelstra{\ss}e 2, 30167 Hannover, Germany}




\section{Supplementary Methods}

\subsection{From topological robustness to error correction}

\label{sec:robust}

Let us describe how to perform a generalized construction of error
correcting models for topologically ordered phases. We start by
considering a topologically ordered phase having degenerate ground
states $\{\ket{g_\alpha}\}$, where any quasilocal operator $V$
satisfies
\begin{equation}
  \bra{g_\alpha} V \ket{g_\beta} = v\delta_{\alpha\beta} + c,
\end{equation}
with $c$ being either zero or vanishing in the thermodynamic limit
\cite{Nussinov2009}. Here, the ground state index $\alpha$ defines a
topological quantum number. We can take this topological quantum
number to be the eigenvalue of an operator $O_\text{topo}$ encoding
topological order. For example, in the case of the toric code,
$O_\text{topo}$ is related to the non-trivial loop operators around
the torus. We now separate the Hamiltonian $H$ into a part containing
the operator $O_\text{topo}$ and a remaining part $H_0$, i.e.,
\begin{equation}
  H = H_0 + \lim\limits_{h\to 0} h O_\text{topo}.
  \label{eq:split}
\end{equation}
Since the topological quantum numbers describing a topological phase
are good quantum numbers by construction, we have $[H, O_\text{topo}]
= [H_0, O_\text{topo}]=0$, at least in the thermodynamic
limit. Importantly, $H_0$ has a unique ground state, as the
topological ground state degeneracy of $H$ is given by
$\dim(O_\text{topo})$.

While it is tempting to try to analyze topological order by
considering the operator $O_\text{topo}$ \cite{denNijs1989}, its
inherent nonlocality severly limits the possibility to make general
statements about its properties. For example, arguments related to
spontaneous symmetry breaking do not apply and hence do not allow to
treat $\langle O_\text{topo}\rangle$ as a topological order
parameter. Therefore, we completely neglect the operator
$O_\text{topo}$ in the following and entirely focus our discusion on
the Hamiltonian $H_0$.

As already mentioned, $H_0$ has a unique ground state $\ket{\psi_0}$,
which is also invariant under all (possibly nonlocal) symmetry
transformations of the original Hamiltonian $H$, as the topological
degrees of freedom have been separated off in
Eq.~(\ref{eq:split}). This means that $\ket{\psi_0}$ describes a
quantum paramagnet (for spin systems) or an insulator (for bosons or
fermions). Without loss of generality, we will use the terminology of
spins systems in the following. Crucially, a paramagnet is
adiabatically connected to the ground state $\ket{\psi_p}$ of the
Hamiltonian
\begin{equation}
  H_B = B \sum\limits_\mu O_\mu,
  \label{eq:hb}
\end{equation}
with quasi-local commuting operators $O_\mu$ and $B$ being a constant
describing an effective magnetic field. The operators $O_\mu$ can be
chosen such that their smallest eigenvalue is zero, meaning that the
ground state satisfies $O_\mu \equiv 0$. Here, we use the index $\mu$
to indicate that the degrees of freedom of $H_B$ are defined on a
different lattice than the original model. Note that this does not
imply that $H_0$ can be written in the form of Eq.~(\ref{eq:hb}), as
the fusion and braiding rules concerning \emph{excited} states on top
of $\ket{\psi_p}$ represent highly nontrivial interactions terms that
are absent in Eq.~(\ref{eq:hb}).

In the following, we choose the state $\ket{\psi_r} =
\ket{\psi_p}\ket{\alpha}$ to serve as the reference state. Which value
of $\alpha$ is chosen to define the reference state is actually not
important as it will not affect any of the $O_\mu$
operators. Importantly, errors with respect to the reference state are
described by violations of the constraint $O_\mu=0$. Since the
topological phase is protected by the gap of the paramagnet, the
paramagnetic phase of $H_0$ is equivalent to the topologically ordered
phase of $H$.

We note that classifying topological phases in terms of the reference
state $\ket{\psi_r}$ is actually not that different from classifying
conventional Landau symmetry breaking phases in terms of local order
parameters, as each order parameter also defines a set of reference
states that maximize the order parameter. Hence, all that is left to
complete the operational definition of topological order in terms of
its error correction abilities is to describe how to construct the
syndrome operators $O_\mu$ from the reference state and how to perform
the error correction.

Since the error syndrome operators $O_\mu$ are quasilocal, it is
rather straightforward to identify them once the reference state is
defined. In particular one can perform an operator expansion in terms
of quasilocal operators $\mathcal{O}_i$ (defined on the original
lattice) to identify the lattice sites $\mu$ and the associated
operators $O_\mu$ in the space of excitations on top of the reference
state. For example, performing an operator expansion on top of the
ground state of the toric code in terms of Pauli matrices gives rise
to the well-known Ising map of the anyons \cite{Trebst2007}. The set
$\{O_\mu\}$ is the set of syndrome operators that needs to be measured
before the error correction procedure can be carried out.

In the following, we assume that the syndrome operators $O_\mu$ have
been measured $M$ times, yielding a set of $\mathcal{O}_\mu^{(r)}$
measurement results, where $r$ running from 1 to $M$ indicates the
individual measurement run. The classification of the phase then
reduces to a purely classical problem: Which operations need to be
applied, such that the reference state having $O_\mu \equiv 0$ is
reached for a given configuration $\mathcal{O}_\mu^{(r)}$ at fixed
$r$? Since the configurations differ for each value of $r$, this
introduces a statistical element to the error correction
circuit. However, since the problem is classical, the required error
correction circuits can be computed for very large system sizes.

Let us now refer to the circuit depth $n_d$ as a suitable statistical
measure (e.g., the mean or the variance) over all $M$ measurement
results. The required operations can be constructed from the
quasilocal operators $\mathcal{O}_i$ used to define the syndrome
operators, as applying their inverse will map the system back onto the
reference state. Remarkably, these operations actually describe the
fusion rules of the topological phase.

Denoting by $d_H$ the Hamming distance (or higher-dimensional
equivalent) to the reference state, i.e., the number of fusion
processes needed to reach the reference state, we can introduce a
simple error correction strategy. From each site $\mu$ containing an
error, we perform a search of the surroundings of $\mu$ to find other
nearby errors. Whenever we find a configuration that allows an
operation that lowers $d_H$, it gets carried out. Although this error
correction algorithm is not necessarily optimal, it is guaranteed to
result in the desired reference state as $d$ is decreasing
monotonously. For the toric code, this strategy precisely yields the
error correction algorithms described in the Methods section.


\subsection{Many-body perturbation theory in the topological phase}

Let us now turn to a scaling analysis of the circuit depth $n_d$ of
this error correction strategy. In the topologically ordered phase, we
can define all the ground states in terms of a perturbative expansion
on top of the Hamiltonian $H_B$ and its ground state $\ket{\psi_p}$,
respectively \cite{Messiah1961}. Formally, the ground state
of the peturbed Hamiltonian $H = H_B + \lambda V$ can be expressed as
\begin{equation}
  \ket{\psi} = \frac{P}{\bra{\psi_p} P \ket{\psi_p}}\ket{\psi_p}
\end{equation}
using the projector $P = \ketbra{\psi}$ given by
\begin{equation}
  P = \ketbra{\psi_p} + \sum\limits_{k=0}^\infty \lambda^k A_{(k)}
\end{equation}
Here, the operators $A_{(n)}$ have the form
\begin{equation}
  A_{(k)} = \sum\limits_{(k)} = -S_{l_1} V S_{l_2} V \cdots V S_{l_{k+1}},
\end{equation}
with the sum running over all sets of $l_i$ satisfying $l_i \geq 0$
and $\sum_i l_i = k$. The resolvent operators $S_l$ are given by
\begin{equation}
  S_l = \begin{dcases}
    \ketbra{\psi_p} & \text{for}\;l=0\\
    \sum\limits_n\frac{\ketbra{n}}{(E_0-E_n)^l}& \text{for}\;l>0
    \end{dcases},
\end{equation}
where $\ket{n}$ are the excited states of $H_B$ with energies
$E_n$. Importantly, the peturturbation
series is convergent as long as we stay in the topological phase, as
there is no closing of the energy gap. Then, we can truncate the
series at $k_{max}$th order, leaving an error in $n_d$ that is exponentially
small in $k_{max}$, i.e., $n_d < C n_{d,k_{max}}$, where $C$ is a constant chosen
to be independent of the system size $N$ and $n_{d,k_{max}}$ is the circuit
depth of the state truncated at order $k_{max}$. Since this expression does
not involve the system size, it is clear that there exists a
finite-depth error correction circuit. This finite-depth scaling is
also reached by the error correction strategy described above, as it
is bounded by the largest cluster size encountered in the perturbative
expansion, which again is a function of $k_{max}$ and not of the system
size.

\subsection{Maximally random errors in the trivial phase}

Importantly, the perturbative argument discussed above breaks down
once the system is outside the topological phase, e.g., in a trivial
phase, as the perturbation series diverges in this case. To establish
the circuit depth scaling for trivial states, let us turn to the error
correction properties of topological phases. We consider a
topologically trivial product state of all spins pointing in the $x$
direction. Such a state has no bit-flip errors (or higher dimensional
equivalents), meaning the $O_\mu^\text{bit}$ describing such errors
are still zero. On the other hand, the $O_\mu^\text{phase}$ related to
phase flip errors are maximally random. This maximum randomness is
reached if all configurations of the phase error syndrome
$\{O_\mu^\text{phase}\}$ are equally likely.

In the following, we consider the mean circuit depth, i.e.,
\begin{equation}
  \bar{n}_d = \frac{1}{M}\sum\limits_r n_d^{(r)},
\end{equation}
where $n_d^{(r)}$ refers to the circuit depth for a given set of
measurements $\mathcal{O}_\mu^{(r)}$. We then proceed by noting that
$n_d^{(r)}$ can be bounded from below by considering a 2D torus
topology containing a single error type (encoding the aforementioned
phase errors) that can be removed by an $m$-ary fusion process, as
adding more error types, higher dimensionality, or open boundaries
will always increase $n_d^{(r)}$. However, treating different $m$
seperately is necessary as we will see below that there is no choice
of $m$ that leads to a minimal $n_d^{(r)}$ for all error
configurations. For $m=2$, this simplification yields the toric code
model in the limit of an infinite magnetic field.

\begin{figure}[ht]
  \includegraphics[width=4cm]{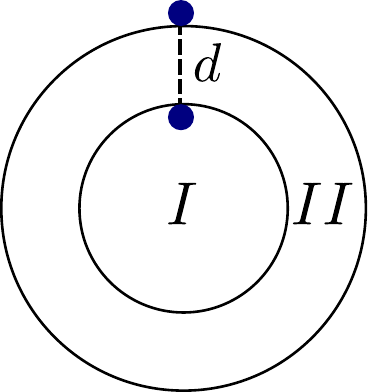}
  \caption{Removal of the last errors. Two parts of the system $I$ and $II$, each containing $N/p$ sites are introduced. With finite probability, the last error in the part $I$ has to be fused across an empty region $II$, requiring at least $d$ steps for the error correction to complete.}
  \label{fig:rings}
\end{figure}
Now, let us show that for the trivial state introduced above, the
circuit depth diverges in the thermodynamic limit. To achieve this, we
look at two neighboring parts within the bulk of the system, one being
a circle and the other a ring surrounding the first part, see
Fig.~\ref{fig:rings}. Each part contains $N/p$ sites, where $p$ is a
constant independent of the system size $N$. To simplify the analysis,
we neglect all error correction steps that occur before we arrive at
less than $m$ errors in the two patches. Obviously, the remaining
steps $n_p^{(r)}$ set again a lower bound on $n_d^{(r)}$. Since all
error configurations are equally likely in the topologically trivial
state under consideration, the remaining number of errors
$N_\varepsilon$ is uniformly distributed between $0$ and $m-1$. Let us
now specialize on the case $N_\varepsilon=1$, which occurs with a
probability of $1/m$. Again, the one remaining error can be located in
either of the two patches with a probability of $1/2$. In the case
where this single error is located in the outer ring, one might get lucky and find the remaining $m-1$ fusion
partners just outside the ring and the remaining circuit depth is
small. However, when the final error is located in the center patch,
finding the fusion partners will require at least
\begin{equation}
  d = \left(\sqrt{2}-1\right)\sqrt{\frac{N}{\pi p}}
\end{equation}
steps as the width $d$ of the empty ring has to be crossed. Since this
configuration occurs with a probability of $1/2m$, the overall circuit
depth has to satisfy
\begin{equation}
  n_d \geq \frac{\sqrt{2}-1}{2m}\sqrt{\frac{N}{\pi p}}
\end{equation}
which diverges in the thermodynamic limit.

Consequently, the circuit depth $n_d$ is always finite in the
topologically ordered phase, while it diverges in the trivial
phase. This demonstrates that the circuit depth can be successfully
used in the classification of topologically ordered phases of matter.

\section{Supplementary Discussion}

\subsection{Error correction for the cubic code}
\begin{figure}[h!]
\begin{center}
\begin{tikzpicture}

\draw (0,0,0) -- (0,0,2) -- (2,0,2) -- (2,0,0) -- cycle;
\draw (0,0,0) -- (0,2,0) -- (2,2,0) -- (2,0,0) -- cycle;
\draw (0,0,0) -- (0,2,0) -- (0,2,2) -- (0,0,2) -- cycle;
\draw (2,0,0) -- (2,2,0) -- (2,2,2) -- (2,0,2) -- cycle;
\draw (0,0,2) -- (0,2,2) -- (2,2,2) -- (2,0,2) -- cycle;
\draw (0,2,0) -- (0,2,2) -- (2,2,2) -- (2,2,0) -- cycle;

\node at (0.2,0.2,0) {$\sigma_0 \otimes \sigma_0$};
\node at (2.15,0.2,0) {$\sigma_0 \otimes \sigma_z$};
\node at (2.2,2.2,0) {$\sigma_z \otimes \sigma_0$};
\node at (0.2,2.2,0) {$\sigma_0 \otimes \sigma_z$};
\node at (0.7,0.2,3.2) {$\sigma_0 \otimes \sigma_z$};
\node at (2.7,0.2,3.2) {$\sigma_z \otimes \sigma_0$};
\node at (2.6,2.2,3.2) {$\sigma_z \otimes \sigma_z$};
\node at (0.6,2.2,3.2) {$\sigma_z \otimes \sigma_0$};
\node at (0.35,-2.5) {$(a)$};
\end{tikzpicture}
\hspace{2.5cm}
\begin{tikzpicture}

\draw (0,0,0) -- (0,0,2) -- (2,0,2) -- (2,0,0) -- cycle;
\draw (0,0,0) -- (0,2,0) -- (2,2,0) -- (2,0,0) -- cycle;
\draw (0,0,0) -- (0,2,0) -- (0,2,2) -- (0,0,2) -- cycle;
\draw (2,0,0) -- (2,2,0) -- (2,2,2) -- (2,0,2) -- cycle;
\draw (0,0,2) -- (0,2,2) -- (2,2,2) -- (2,0,2) -- cycle;
\draw (0,2,0) -- (0,2,2) -- (2,2,2) -- (2,2,0) -- cycle;

\node at (0.2,0.2,0) {$\sigma_x \otimes \sigma_x$};
\node at (2.15,0.2,0) {$\sigma_0 \otimes \sigma_x$};
\node at (2.2,2.2,0) {$\sigma_x \otimes \sigma_0$};
\node at (0.2,2.2,0) {$\sigma_0 \otimes \sigma_x$};
\node at (0.7,0.2,3.2) {$\sigma_0 \otimes \sigma_x$};
\node at (2.6,0.2,3.2) {$\sigma_x \otimes \sigma_0$};
\node at (2.6,2.2,3.2) {$\sigma_0 \otimes \sigma_0$};
\node at (0.6,2.2,3.2) {$\sigma_x \otimes \sigma_0$};
\node at (0.35,-2.5) {$(b)$};
\end{tikzpicture}
\end{center}
\caption{Sketch of the cubic code model. Decomposition of the operators $A_c$ (a) and $B_c$ (b) in terms of Pauli matrices $\sigma_x$ and $\sigma_z$, with $\sigma_0$ being the identity.}
\label{fig:cubic}
\end{figure}
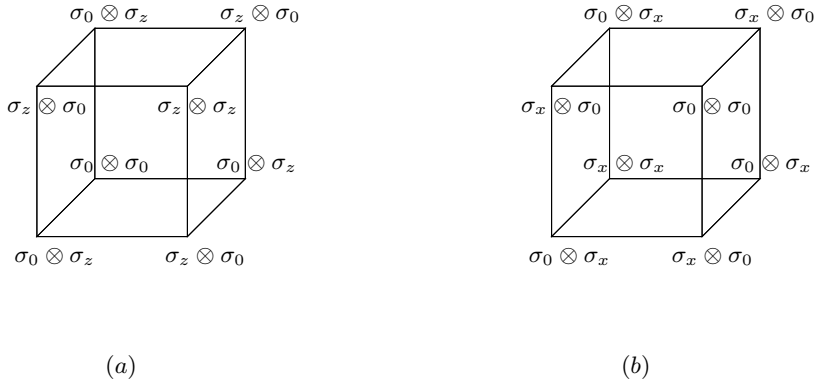
  
\begin{figure}[ht]
  \begin{tabular}{cp{0.5cm}c}
    \includegraphics[width=6cm]{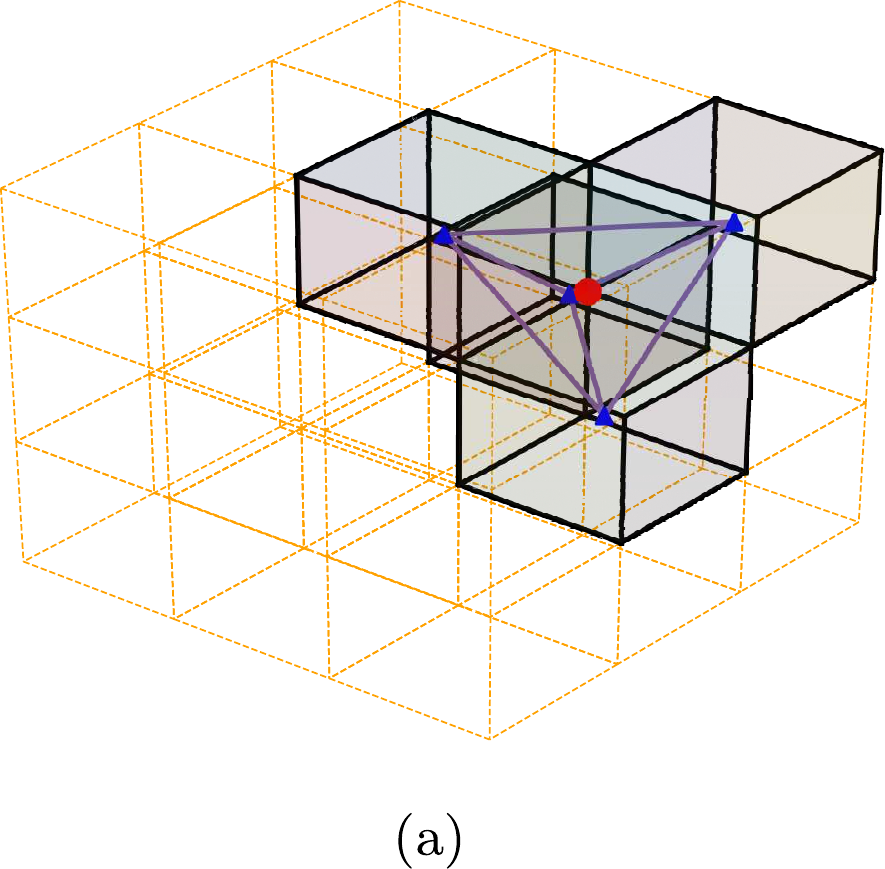} && \includegraphics[width=6cm]{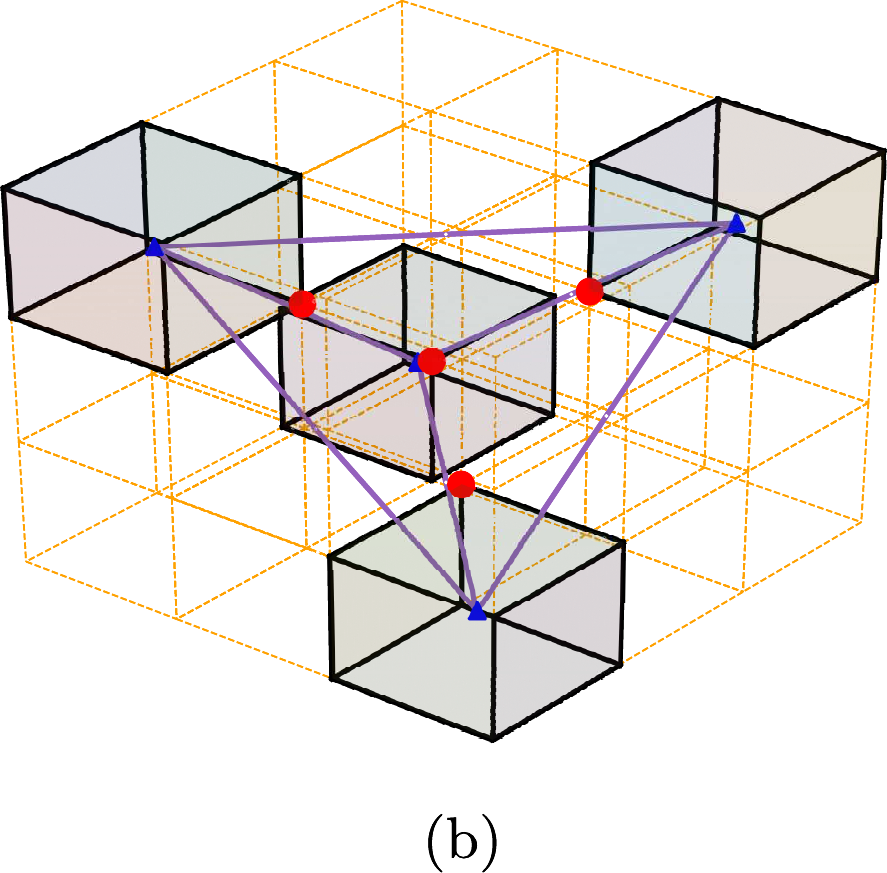}
  \end{tabular}
    \caption{Errors of the cubic code. (a) Applying a single spin-flip operator (red) to the ground state creates four errors (blue) shaped in a tetrahedron form. (b) The errors can be moved around by increasing the size of the tetrahedron by applying three additional spin-flip operators.}
    \label{fig:tetra}
\end{figure}

The cubic code \cite{Haah2011} is a paradigmatic model for fracton
excitations, i.e., the fusion of the errors in no longer described by
a linear string operator as in the toric code, but an operator that
involves the sites inside the volume spanned by the errors in a
fractal shape. The Hamiltonian of the cubic code has the same structure as the toric code, i.e.,
\begin{equation}
  H = -E_0 \sum_c (A_c + B_c).
  \label{eq:Hcubic}
\end{equation}
The crucial properties of the cubic code arise from the definition of
the $A_c$ and $B_c$ operators. As shown in Fig.~\ref{fig:cubic}, both
operators are defined in terms of the cubes $c$, with each vertex of
the cube supporting two spin-1/2 lattice sites. Owing to the
arrangement of the Pauli matrices $\sigma_x$ and $\sigma_z$, flipping
a single spin will result in the appearance of four errors arranged in
a tetrahedron, see Fig.~\ref{fig:tetra}. These errors can no longer be
moved around by additional flips of single spins, as such an operation
would remove one error and create three additional ones, i.e., changing
the number of errors and thus the energy of the system. Instead,
moving errors require the application of an operator that involves the
spins inside the tetrahedron in a fractal shape.

Nevertheless, error correction in the cubic code can be implemented in
the usual way, with the reference state being one of the ground states
of Eq.~(\ref{eq:Hcubic}). Here, each walker has to search for the
existence of two other errors located at the vertices of a
tetrahedron. If such triples of errors can be found, fusion using the
fractal operator will lower the Hamming distance to the reference
state, even if the fourth vertex of the tetrahedron does not contain
an error.

Having specified the error correction algorithm, we can also look into
the consequences for the circuit depth. In the fracton phase, the
perturbative argument introduced in Sec.~\ref{sec:robust} holds and
the circuit depth is finite. For the trivial phase, we consider the
case where all error syndroms for one error type (e.g., $A_c$) are
equally likely. In this case, there are again configurations occuring
with finite probability that require the application of a
thermodynamically large fractal operator (i.e., divering with
$N$). From this, we conclude that we can also successfully classify
fracton phases using our error correction approach.

\clearpage
